\documentstyle[aps,multicol,epsf]{revtex}

\begin{document}

\title{
Renormalization group for evolving networks
}

\author{
S.N. Dorogovtsev$^{1, 2, \ast}$  
}

\address{
$^{1}$ A.F. Ioffe Physico-Technical Institute, 194021 St. Petersburg, Russia\\
$^{2}$ Departamento de F\'\i sica and Centro de F\'\i sica do Porto, Faculdade 
de Ci\^encias, 
Universidade do Porto,\\
Rua do Campo Alegre 687, 4169-007 Porto, Portugal
}

\maketitle

\begin{abstract} 
We propose a renormalization group treatment of stochastically growing networks. As an example, we study percolation on growing scale-free networks in the framework of a real-space renormalization group approach. 
As a result, we find that the critical behavior of percolation on the growing networks differs from that in uncorrelated nets. 
\end{abstract}

\pacs{05.10.-a, 05-40.-a, 05-50.+q, 87.18.Sn}

\begin{multicols}{2}

\narrowtext


Evolving networks with a complex distribution of connections attract much interest from a wide circle 
of researchers \cite{w99,s01,ab02,dm02,bookdm02}. The physicists' contribution to the rapid progress in this field is based on understanding that networks are objects of classical statistical mechanics and can be effectively studied by using standard approaches of statistical physics. 

In this communication, we demonstrate how a real-space renormalization group approach (or positional renormalization group), which is traditional in critical phenomena theory (see, e.g., Refs. \cite{k76,m76,jrw78,ys79,sf80,rb81} and references therein), can be applied to stochastically growing networks. For the demonstration, we consider a bond percolation problem, although other models of cooperative behavior on evolving networks can be investigated in a similar way. 

Percolation on uncorrelated equilibrium networks is well studied 
\cite{cebh00,cnsw00,nsw00,cbh02}. In fact, many other cooperative models 
(with discrete symmetry of the order parameter and absence of frustrations) on equilibrium networks   
(the spread of diseases \cite{pv01,nmej02}, the Ising and the Potts models \cite{isingetc02}, etc.) show a behavior similar to that for the percolation. In this respect, percolation problems are very representative.   

Percolation on a growing network is a more complex problem than for equilibrium ones.   
The reason for this complexity is a wide spectrum of correlations, which are inevitably present in growing nets. Note that the correlations between the degrees of the nearest-neighbor vertices \cite{kr01,pvv01,ms02}, which were thoroughly studied in equilibrium networks \cite{nmejb02,bl02,bpv02,vm02}, is only a particular kind of the correlations.  

\begin{figure}
\epsfxsize=70mm
\centerline{\epsffile{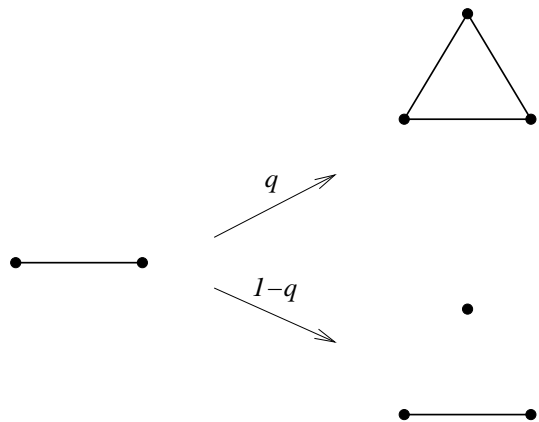}} 
\caption{
Edge transformation which generates the stochastically growing scale-free network. At each time step, each edge of the network transforms into one of the two shown configurations with the complementary probabilities $q$ and $1-q$.     
}
\label{f1}
\end{figure}

We study the following problem. First we grow an infinite size network and then consider a classical bond percolation problem on it. That is, randomly chosen edges of the infinite network are simultaneously removed. A fraction $p$ of edges is retained. We use the model of a stochastically growing, undirected, highly-clustered, scale-free network \cite{bookdm02} (see Fig. \ref{f1}), which is ideally suited for a real-space renormalization group procedure. 
In principle, the percolation problem for this network can be exactly solved, at least, at some particular case. We, however, use this model for the first demonstration of the renormalization group method for growing random networks. 

The second moment of the degree distribution $P(k)$ of this growing network diverges (degree is the total number of connections of a vertex, sometimes it is called ``connectivity''). We show that the ``percolation threshold'' is zero, $p_c=0$, that is the percolating cluster (the giant connected component) exists at each $p \neq 0$ (super-stability against random damage). The $\gamma$ exponent of the degree distribution $P(k) \propto k^{-\gamma}$ of the net is below $3$. We find that all the derivatives of the relative size of the giant component $M(p)$ over $p$ diverge at the percolation threshold. Moreover, $M(p \ll 1) \sim e^{-\text{const}/p}$.  This differs from the corresponding critical singularity for percolation on uncorrelated scale-free networks with exponent $\gamma < 3$. 
     
The network is constructed in the following way \cite{bookdm02}. 
The growth starts from a single edge connecting two vertices ($t=0$). 
At each time step, each edge of the network transforms as shown in Fig. \ref{f1}:

\begin{list}{}{\leftmargin=17pt} 

\item[(a)]
with probability $q$, an edge ``creates'' a vertex which is attached to both the end vertices of its ``mother'' edge, or  

\item[(b)]
with the complementary probability $1-q$, this edge creates a bare vertex. 

\end{list} 


\noindent
Of course, we can equally leave an edge unchanged in item (b), 
and, at first sight, the creation of bare vertices seems superfluous. However, this inessential feature of the transformation will be convenient for us.  

In the particular case $q=1$, this network turns out to be the simple deterministic graph that was introduced in Ref. \cite{dm02} and studied in detail in Ref. \cite{dgm02}. This graph ($q=1$) belongs to a wide class of deterministic growing nets which are currently being intensively studied \cite{brv01,jkk02,n02}. 
If $q<1$, the growth is stochastic. As $q \to 0$ (but $q \neq 0$) the network coincides with a network growing by attaching a new vertex to the ends of a randomly chosen edge at each time step \cite{dms01}. The latter has practically the same degree distribution as the Barab\'asi-Albert model ($\gamma = 3$) but is strongly clustered. 
Indeed, the network that is generated by the transformation shown in Fig. \ref{f1} has numerous loops of length three. So it has a large clustering coefficient at any value of $q$.    

One can easily modify the transformation in such a way that the resulting network will turn out to be exactly the Barab\'asi-Albert model in the limiting case $q \to 0$ ($q \neq 0$). Let at each time step, each edge creates a new vertex which, with probability $q$, is attached to one end vertex of the edge and, with the same probability, is attached to the other end vertex. At $q=1$, we again obtain the deterministic graph \cite{dm02,dgm02}, and as $q \to 0$, the network approaches the Barab\'asi-Albert model with zero clustering.  

We, however, apply the transformation that provides the strongly clustered networks at each $q$. Note that 
we use transformations generating 
networks with the small-world effect \cite{w99}, that is the average shortest-path length of the network grows logarithmically with the network size.       

Simple calculations show that the degree distribution of the network is scale-free with exponent 
\begin{equation}
\gamma = 1 + \frac{\ln(1+2q)}{\ln(1+q)}
\, .   
\label{1}
\end{equation} 
The spectrum of degrees is continuous at $q < 1$. 
As $q$ decreases from $1$ to $0$, $\gamma$ grows from $1+ \ln3/\ln2=2.585\ldots$ to $3$. The average number of edges in the network grows as $(1+2q)^t$, where $t$ is the number of a time step. 

For demonstration, we use the simplest, rather naive version of the real-space renormalization group approach \cite{k76,m76,jrw78,ys79,rb81,d82}.  
Let us outline the procedure in application to our network.  
We stop the grows at a time step $T \to \infty$ and spoil the network: 
if an edge is present in the undamaged net, then this edge is retained in the damaged net with the probability $p$.   
Then we invert the transformation in Fig. \ref{f1} and define $n = T-t$ for the inverted transformation, which is actually a decimation procedure. Further, we introduce the probability $p_n$ that if an edge connects a pair of vertices of the undamaged net at $t=T-n$, then at the $n$-th step of the decimation for the damaged net, there is 
connection
 between these vertices. Here, $p_0=p$. One can easily derive the following recursion relation for $p_n$:    
\begin{eqnarray}
p_{n+1} = 
& & q [p_n^3 + 3 p_n^2(1-p_n) + p_n(1-p_n)^2] 
\nonumber
\\[5pt]
& & 
+ (1-q)p_n
\, .    
\label{2}
\end{eqnarray} 
Its structure is evident from Fig. \ref{f1}. 

Let us find the dependence of the relative size $M$ of the percolating 
cluster on $p$. Here, for convenience, we define $M$ as the fraction of {\em edges} that belong to the giant connected component of the network (the percolating cluster). So, $M(p=1)=1$ in the undamaged network. 
As is usual, for calculating $M$, we find the average number $m_n$ of edges in the ``percolation'' configurations of the renormalization group transformation at an $n$-th step: 
\begin{eqnarray}
& & 
p_{n} m_{n} = 
\nonumber
\\[5pt]
& &   
q [3\cdot p_{n-1}^3 + 2\cdot3 p_{n-1}^2(1-p_{n-1}) + 1\cdot p_{n-1}(1-p_{n-1})^2] 
\nonumber
\\[5pt]
& & 
+ (1-q)p_{n-1} 
\,    
\label{3}
\end{eqnarray} 
[compare Eqs. (\ref{2}) and (\ref{3}); notice also the multiple $p_n$ on the left-hand side of Eq. (\ref{3})]. 
Then, $M(p)$ can be obtained from the following relation 
\begin{equation}
M = \prod^\infty_{n=1} \frac{m_{n}}{1+2q}
\, ,   
\label{4}
\end{equation} 
where we have taken into account that the number of edges in the undamaged network increases by $1+2q$ times at each time step of the evolution.  

The final forms of the recursion relation for $p_n$ and the expression for the relative size of the percolating cluster are   
\begin{eqnarray}
& & p_{n+1} = p_n[q p_n (1-p_n) + 1]
\, ,
\nonumber
\\[5pt] 
& & M(p) = p \prod^\infty_{n=1} \frac{1 + 2qp_n(2-p_n)}{1+2q}
\, ,  
\label{5}
\end{eqnarray}
where $p_0 = p$. 
One can see that the recursion relation for $p_n$ has only two fixed points, $0$ and $1$. At any $p_0 = p \neq 0$, $p_n$ approaches $1$ as $n \to \infty$, 
which indicates the presence of the percolating cluster. So, the percolating cluster of this network cannot be eliminated by the random removal of edges at any value of the parameter $q$. 

\begin{figure}
\epsfxsize=79mm
\centerline{\epsffile{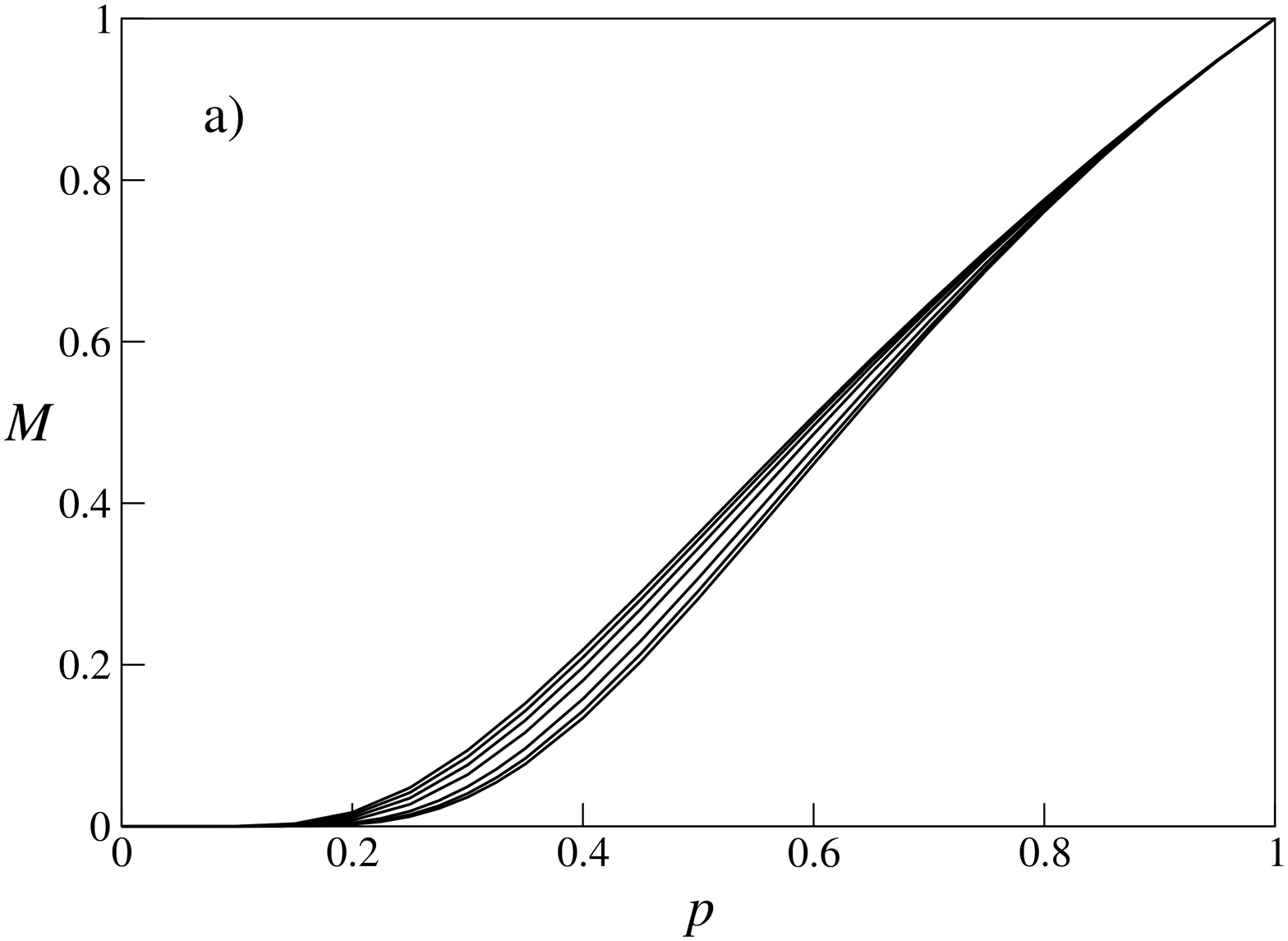}}
\epsfxsize=79mm
\centerline{\epsffile{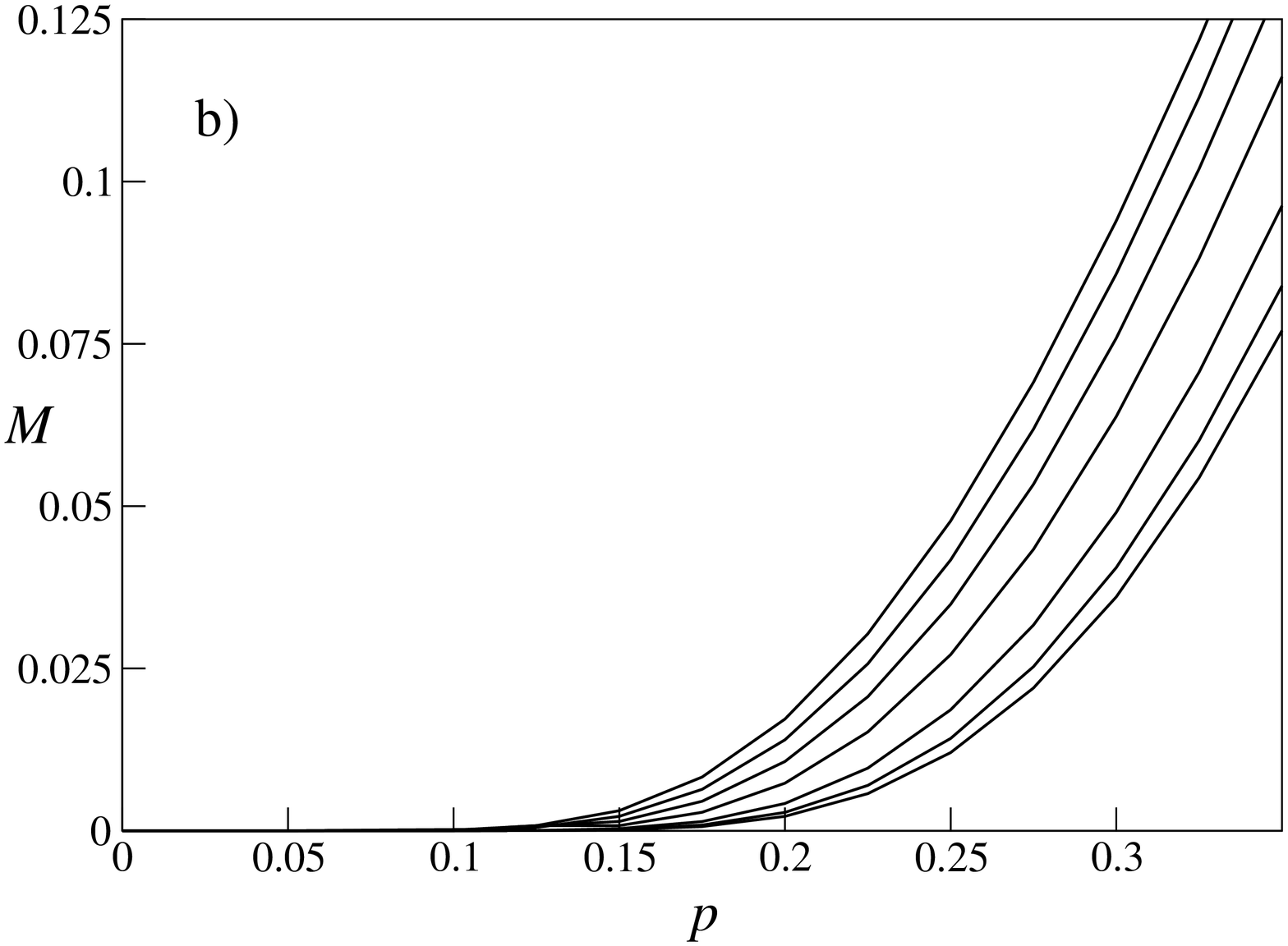}} 
\caption{
(a) Relative size of the percolating cluster, $M$, as the function of the fraction of retained edges, $p$. 
The curves correspond to the values of the parameter $q$: 1.0, 0.8, 0.6, 0.4, 0.2, 0.1, 0.05, from top to bottom.  
(b) The same as (a) but in the region of small $p$.    
}
\label{f2}
\end{figure}

By using the relations (\ref{5}), we numerically obtain the dependence $M(p)$ 
(see Fig. \ref{f2}). From Fig. \ref{f2}(a), one can see that the curves $M(p)$ weakly depend on the parameter $q$ if $p$ is large enough. 
The essential difference is visible in the range of small $p$, see Fig. \ref{f2}(b).  
The analytical analysis of the relations (\ref{5}) immediately yields the asymptotic  
behavior of $M(p)$ at small $p$:     
\begin{equation}
M(p \ll 1) \sim \, \exp\left[ -\frac{\ln(1+2q)}{q}\, p^{-1}\right]
\, .   
\label{6}
\end{equation} 
We do not write down the preexponential factor, since our simplified, demonstrative renormalization group procedure certainly cannot give its proper form. In fact, at each step of the procedure, the complex spectrum of the probabilities of the realization of various edge configurations is renormalized to (one can also say, is substituted by) a single delta-function distribution. 
If a cooperative model under consideration does not contain frustrations, 
then usually, this substitution is not dangerous (we discuss only qualitative results). This is the case for percolation.  
Only for $q \ll 1$, this approximation can cause serious problems [e.g., very inaccurate values of the resulting factor in the exponential 
of Eq. (\ref{6})]. 


Expression (\ref{6}) shows that all the derivatives of $M(p)$ diverge at the ``percolation threshold'' $p_c = 0$. The type of this divergence differs from that for uncorrelated scale-free networks with $2 < \gamma < 3$, where 
$M(p \ll 1) \sim p^{1/(3-\gamma)}$ and only if $\gamma=3$, the relative size of the percolating cluster behaves as $M(p \ll 1) \sim e^{-\text{const}/p}$ (see Ref. \cite{cbh02}). 

There are two possible reasons for this difference. The first is the wide spectrum of correlations induced by the growth. The second is the high clustering and numerous loops in our network. The loops, in principle, may lead to fluctuation effects. (These effects are absent in cooperative models on uncorrelated networks, which have a tree-like local structure.) 

One has to admit that a real-space renormalization group approach has a reputation of an uncontrolled approximation. Moreover, in our demonstration, we have used the simplest version of the approach, which has allowed us to obtain analytical results. 
In principle, one can use more refined versions of the real-space renormalization group procedure (e.g., see Ref. \cite{sf80}), where, however, numerical calculations are necessary \cite{remark}.      

We considered one of traditional cooperative models on undirected growing networks. However, the renormalization group approach can also be used as a tool for studying structural properties (e.g., a distribution of loops \cite{bc02}, etc.) of evolving networks. Directed networks can be considered in a similar way as for randomly directed percolation on a lattice \cite{rb81}. 
Furthermore, the transformation that we used for generating the growing random network (see Fig. \ref{f1}) is only a simple example and can be easily generalized  \cite{bookdm02}.    

In this communication, the renormalization group approach was applied to evolving networks. One should mention the applications of a real-space renormalization group technique to static networks: to small-world networks in Ref. \cite{nw99} and, recently, to a random network of masses connected by springs in Ref. \cite{h02}. 

In summary, we have applied the real-space renormalization group procedure to stochastically growing networks. In the framework of this approach, we have considered the percolation problem for scale-free growing networks with exponent $\gamma$ in the interesting region $\gamma < 3$. 
The percolation threshold is zero, $p_c = 0$, but the observed critical behavior essentially differs from that for uncorrelated networks. Our results demonstrate new possibilities of a renormalization group method.  
\\[-5pt]

This work was  
partially  
supported by the project POCTI/1999/FIS/33141. 
The author thanks A.V.~Goltsev and A.N.~Samukhin for many useful discussions.  
Special thanks to the Centro de F\'\i sica do Porto and J.F.F.~Mendes.  
\\

\noindent
{\small $^{\ast}$      Electronic address: sdorogov@fc.up.pt} 

\end{multicols} 


\begin{references}  


\bibitem{w99} D.J. Watts,  
{\it Small Worlds: The Dynamics of Networks between Order and Randomness} 
(Princeton University Press, Princeton, NJ, 1999). 

\bibitem{s01} 
S.H. Strogatz, Nature {\bf 401}, 268 (2001). 

\bibitem{ab02} 
R. Albert and A.-L. Barab\'asi, Rev. Mod. Phys. {\bf 74}, 47 (2002)). 

\bibitem{dm02}
S.N. Dorogovtsev and J.F.F. Mendes, 
Adv. Phys. {\bf 51}, 1079 (2002). 
 
\bibitem{bookdm02}
S.N. Dorogovtsev and J.F.F. Mendes, 
{\em Evolution of Networks: From Biological Nets to the Internet and WWW} 
(Oxford University Press, Oxford, 2003).    

\bibitem{k76}  
L.P. Kadanoff, Ann. Phys. (N.Y.) {\bf 100}, 359 (1976). 

\bibitem{m76} 
A.A. Migdal, Sov. Phys. JETP {\bf 42}, 413 (1976); {\bf 42}, 743 (1976).  

\bibitem{jrw78} 
C. Jayaprakash, E.K. Riedel, and M. Wortis, 
Phys. Rev. B {\bf 18}, 2244 (1978). 

\bibitem{ys79} 
J.M. Yeomans and R.B. Stinchcombe, J. Phys. C {\bf 12}, 347 (1979). 

\bibitem{sf80}  
M. Schwartz and S. Fishman, Physica A {\bf 104}, 115 (1980); 
A. Efrat and M. Schwartz, cond-mat/0212602. 

\bibitem{rb81}
S. Redner, J. Phys. A {\bf 14}, L349 (1981); 
S. Redner and A.C. Brown, J. Phys. A {\bf 14}, L285 (1981). 

\bibitem{cebh00} R. Cohen, K. Erez, D. ben-Avraham, and S. Havlin, 
Phys. Rev. Lett. {\bf 85}, 4626 (2000); 
Phys. Rev. Lett {\bf 86}, 3682 (2001).  

\bibitem{cnsw00} 
D.S. Callaway, M.E.J. Newman, S.H. Strogatz, and D.J. Watts,
Phys. Rev. Lett. {\bf 85}, 5468 (2000). 

\bibitem{nsw00} 
M.E.J. Newman, S.H. Strogatz, and D.J. Watts, 
Phys. Rev. E {\bf 64}, 026118 (2001). 

\bibitem{cbh02}
R. Cohen, D. ben-Avraham, and S. Havlin, 
Phys. Rev. E {\bf 66}, 036113 (2002). 

\bibitem{pv01} R. Pastor-Satorras and A. Vespignani, 
Phys. Rev. Lett. {\bf 86}, 3200 (2001);  
Phys. Rev. E {\bf 63}, 066117 (2001). 

\bibitem{nmej02} 
M.E.J. Newman, Phys. Rev. E {\bf 66}, 016128 (2002). 

\bibitem{isingetc02} 
A. Aleksiejuk, J.A. Holyst, and D. Stauffer, Physica A {\bf 310}, 260 (2002);
S.N. Dorogovtsev, A.V. Goltsev, and J.F.F. Mendes, Phys. Rev. E {\bf 66}, 016104 (2002); 
M. Leone, A. V\'azquez, A. Vespignani, and R. Zecchina, 
Eur. Phys. J. B {\bf 28}, 191 (2002);    
A.V. Goltsev, S.N. Dorogovtsev, and J.F.F. Mendes, cond-mat/0204596; 
F. Igl\'oi and L. Turban, Phys. Rev. E {\bf 66}, 036140 (2002).  

\bibitem{kr01} 
P.L. Krapivsky and S. Redner, Phys. Rev. E {\bf 63}, 066123 (2001). 

\bibitem{pvv01}
R. Pastor-Satorras, A. V\'azquez, and A. Vespignani, 
Phys. Rev. Lett. {\bf 87}, 258701 (2001). 

\bibitem{ms02}
S. Maslov and K. Sneppen, 
Science, {\bf 296}, 910 (2002). 

\bibitem{nmejb02}
M.E.J. Newman, Phys. Rev. Lett. {\bf 89}, 208701 (2002). 

\bibitem{bl02} 
J. Berg and M. L\"assig, Phys. Rev. Lett. {\bf 89}, 228701 (2002). 

\bibitem{bpv02}
M. Bogu\~n\'a, R. Pastor-Satorras, and A. Vespignani, 
cond-mat/0208163. 

\bibitem{vm02}
A. V\'azquez and Y. Moreno, cond-mat/0209182.

\bibitem{dgm02} 
S.N. Dorogovtsev, A.V. Goltsev, and J.F.F. Mendes, 
Phys. Rev. E {\bf 65}, 066122 (2002). 

\bibitem{brv01}
A.-L. Barab\'asi, E. Ravasz, and T. Vicsek, 
Physica A {\bf 299}, 559 (2001). 

\bibitem{jkk02}  S. Jung, S. Kim, and B. Kahng, 
Phys. Rev. E {\bf 65}, 056101 (2002). 

\bibitem{n02}  J.D. Noh, cond-mat/0211399. 

\bibitem{dms01}
S.N. Dorogovtsev, J.F.F. Mendes, and A.N. Samukhin, 
Phys. Rev. E {\bf 63}, 062101 (2001). 

\bibitem{ba99} A.-L. Barab\'asi and R. Albert, 
Science {\bf 286}, 509 (1999).

\bibitem{d82}
S.N. Dorogovtsev, Sov. Phys. Solid State, {\bf 24}, 55 (1982); 
Sov. Phys. Solid State, {\bf 24}, 948 (1982).











\bibitem{remark} Note another week point of our simple procedure. We actually assumed that 
if the percolating cluster is present in the network, then 
``the oldest'', highly connected vertices belong to this cluster. In principle, this may not be the case if $q$ is small (and $p \to 0$). 
This is the second reason why our results are only qualitative, at least, in the region of small $q$, 
where exponent $\gamma$ is close to $3$.  

\bibitem{bc02}  G. Bianconi and A. Capocci, 
cond-mat/0212028. 

\bibitem{nw99}  M.E.J. Newman and D.J. Watts,  
Phys. Lett. A {\bf 263}, 341 (1999). 

\bibitem{h02}  M.B. Hastings,  
cond-mat/0212303. 


\end{references}
\end{document}